\documentclass[pra,twocolumn,showpacs]{revtex4-2}
\usepackage{amsmath,amscd,amssymb,color}
\usepackage{graphicx,amsfonts,dsfont}
\usepackage{braket}
\usepackage{epstopdf}
\usepackage{hyperref}
\usepackage{float}
\usepackage{enumerate,bbold}
\usepackage{array}
\newcommand{\ie}{\textit{i.e.}}
\begin{document}
\title{Experimental demonstration of the dynamics of quantum
coherence evolving under a PT-symmetric Hamiltonian on an
NMR quantum processor}
\author{Akanksha Gautam}
\email{akankshagautam@iisermohali.ac.in}
\affiliation{Department of Physical Sciences, Indian
Institute of Science Education \& 
Research Mohali, Sector 81 SAS Nagar, 
Manauli PO 140306 Punjab India.}
\author{Kavita Dorai}
\email{kavita@iisermohali.ac.in}
\affiliation{Department of Physical Sciences, Indian
Institute of Science Education \& 
Research Mohali, Sector 81 SAS Nagar, 
Manauli PO 140306 Punjab India.}
\author{Arvind}
\email{arvind@iisermohali.ac.in}
\affiliation{Department of Physical Sciences, Indian
Institute of Science Education \& 
Research Mohali, Sector 81 SAS Nagar, 
Manauli PO 140306 Punjab India.}
\affiliation{Vice Chancellor, Punjabi University Patiala,
147002, Punjab, India}
\begin{abstract}
In this work, we study the dynamics of quantum coherence (total coherence,
global coherence and local coherence) evolving under a local
PT-symmetric Hamiltonian in maximally entangled bipartite and
tripartite states. Our results indicate that quantum coherence in the
bipartite state oscillates in the unbroken phase regime of the
PT-symmetric Hamiltonian. Interestingly, in the broken phase regime,
while the global coherence decays exponentially, the local
and total coherences enter a ``freezing'' regime where
they attain a stable value over time.  A similar pattern is observed
for the dynamics of total and local coherences in the maximally
entangled tripartite state, while the dynamics of global coherence in
this state differs from that of the bipartite state. 
These results were experimentally validated  for a
maximally entangled bipartite state on a three-qubit nuclear magnetic
resonance (NMR) quantum processor, with one of the qubits acting as an
ancilla. The experimental results match well with the theoretical
predictions, upto experimental errors. 
\end{abstract} 
\maketitle 
\section{Introduction}
\label{intro} 
The hermiticity of physical observables in standard quantum mechanics ensures
both real energy spectra of quantum systems and unitary evolution of closed
systems. In 1998, Bender and Boettcher discovered certain parity-time (PT)
symmetric Hamiltonians that despite being non-hermitian, can still have real
energy spectra, thereby demonstrating that hermiticity is a sufficient condition
but not necessary~\citep{Bender-prl-98,Bender-iop-2007}. It was observed that
that the eigenvalues of the PT-symmetric Hamiltonian are always real and
PT-symmetry is unbroken if the eigenstates of a PT-symmetric Hamiltonian are
also eigenstates of the PT operator; otherwise, the eigenvalues are complex and
PT-symmetry is broken. The point where the eigenvalues and eigenvectors
coalesce, causing a transition from unbroken to
broken PT-symmetry is called the exceptional point\citep{ozdemir-natMat-2019,Ganainy-natPhy-2018}.

PT-symmetric Hamiltonians have been implemented on various experimental
platforms including ultracold atoms \citep{Li-natcomm-2019}, superconducting
circuits \citep{Quijand-pra-2018,Naghiloo-natphy-2019}, nitrogen-vacancy
centers in diamonds \citep{Pick-prr-2019,Wu-sci-2019}, optical waveguides
\citep{Ruter-natphy-2010,Klauck-natphot-2019} and NMR spins
\citep{Wen-pra-2019}. The anti-PT-symmetric Hamiltonian was
implemented using NMR spins where information flow between 
the system and
environment was experimentally explored~\cite{Wen-npj-2020}.

PT-symmetric Hamiltonians have attracted wide attention due to their
interesting features, and numerous studies have been performed including
increase or restoration of entanglement by a local PT-symmetric Hamiltonian
\citep{Wen-pra-2019,Chen-pra-2014}, protecting quantum correlations via
non-hermitian operations \citep{Wang-qip-2018}, no-signaling principle violation
\citep{lee-prl-2014,Tang-natphot-2016}, observing fast evolution in a
PT-symmetric system \citep{Gunther-prl-2004,Bender-prl-2007} and observing
critical phenomena in a PT-symmetric system
\citep{kawabata-prl-2017,Xiao-prl-2019}. Recently, it has been shown that
quantum coherence is increased in a single-qubit system under PT-symmetric
Hamiltonian and maximal quantum coherence is observed at the exceptional
point\citep{Naikoo-iop-2021,Wang-pra-2021}. The flow of quantum coherence of a
single qubit under PT-symmetric and anti-PT-symmetric Hamiltonians has been
demonstrated on an optics system \cite{Fang-commphy-2021}.

Different measures of basis dependent quantum coherence have been proposed such
as $l_{1}$-norm and relative entropy 
~\citep{Baugratz-prl-2014,rastegin-pra-2016,rana-pra-2016,Shao-pra-2015}. Based
on these measures, various types of quantum coherences have been defined in
multipartite systems such as global coherence, local coherence and total
coherence~\citep{Cao-pra-2020,Radha-prl-2016}.  Recently, bipartite and
tripartite entanglement has been generated in a PT-symmetric system that
involves three interacting cavities \citep{Leduc-symm-2021}. The evolution
process of entropy and entanglement in a three-qubit system has been
investigated using nuclear spins \citep{Wen-prr-2021}.

In this work, we use the relative entropy of
coherence~\citep{Baugratz-prl-2014} to quantify quantum coherence of various
kinds present in a multipartite system, and explicitly study their dynamics
in a bipartite and a tripartite system evolving under a local PT-symmetric
Hamiltonian.  In the maximally entangled bipartite (Bell) and tripartite (GHZ)
states, only global coherence exists while local coherence remains zero.  Our
theoretical results show that local coherence, which previously did not exist
in such maximally entangled states, gets created under the local PT-symmetric
Hamiltonian.  Global, local and total coherences in the bipartite entangled
state show an oscillatory behavior in 
the unbroken PT-symmetry regime.  On the
other hand, in the broken PT-symmetry regime, global coherence first
increases then decays exponentially; total coherence also increases initially
but  attains a stable value at later times, while
local coherence is first created, then increases and
stabilizes at later times.  Similar patterns of coherence
dynamics is observed in the tripartite maximally entangled state, in the
unbroken regime of PT-symmetry. In the broken phase of PT-symmetry, the pattern
of total and local coherences are similar to the bipartite state but the global
coherence exhibits a different dynamics. Further, the dynamics of quantum
coherence present in a two-qubit reduced state of a tripartite state was
investigated under the PT-symmetric Hamiltonian in 
both unbroken and broken phase
regimes.  We have also  experimentally demonstrated the dynamics of quantum
coherence of a maximally entangled bipartite state using three NMR qubits.  Two
of the qubits were utilized to prepare two-qubit states and the third qubit was
used as an ancillary qubit to simulate a PT-symmetric Hamiltonian acting on one
of the qubits. Both broken as well as unbroken phase regimes of the
PT-symmetric Hamiltonian were simulated experimentally and the dynamics of
various types of of quantum coherences in the maximally entangled bipartite
state was verified.

This article is organized as follows: Section \ref{simu} describes the
procedure to simulate a general PT-symmetric Hamiltonian. Section \ref{dyna}
contains details of the simulation of the dynamics of quantum coherence (global
coherence, local coherence and total coherence) in a maximally entangled
bipartite (Bell) state and a tripartite (GHZ) state evolving under the local
PT-symmetric Hamiltonian. Section \ref{Exp} contains the experimental results
of the implementation of the local PT-symmetric Hamiltonian on a maximally
entangled bipartite state. Section \ref{Con} contains some concluding remarks.

\section{Simulation of the PT-Symmetric Hamiltonian}
\label{simu}
\begin{figure}[t]
\includegraphics[angle=0,scale=1]{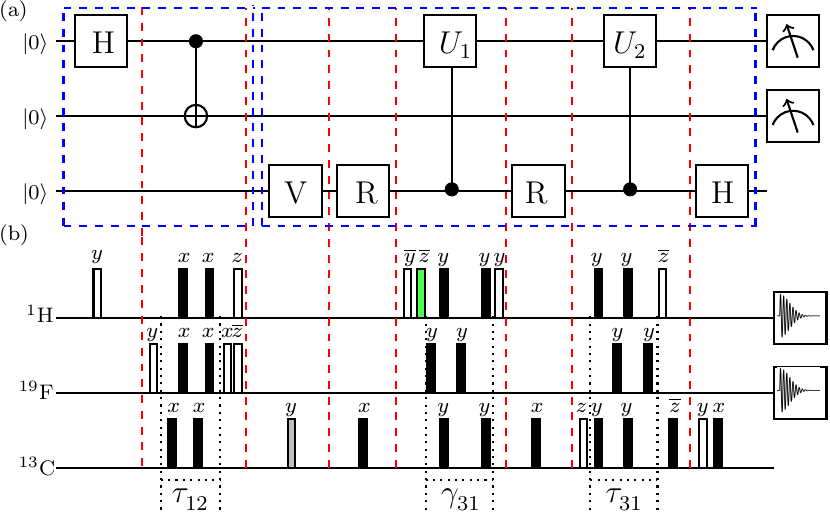}
\caption{(a) Quantum circuit to simulate a PT-symmetric Hamiltonian. The gates
in the first box create a maximally entangled Bell state ($\rho_{_{BS}}\otimes
\rho_{0}$) on the first two qubits, with the third qubit acting as an ancillary
qubit. The action of the $V$, $U_1$ and $U_2$ gates is described in the main
text. $H$ is a Hadamard gate and $R$ acts as a NOT gate when $\phi$ (given in
Eq.\ref{phi}) is negative, otherwise $R$ is an $I_{2}$ identity operation. (b)
The corresponding NMR pulse sequence, where the unfilled rectangles are
$\frac{\pi}{2}$ pulses, the black rectangles denote $\pi$ pulses, the grey
rectangle is a $\theta$ pulse and the green rectangle denotes a $\phi$ pulse.
Pulse phases are given above each pulse and a bar over a phase represents
negative phase. The free evolution time intervals $\tau_{12}$ and $\tau_{31}$
are given by 1/2$J_{12}$ and 1/2$J_{31}$ respectively, where $J_{12}$ and
$J_{31}$ are scalar coupling constants.} 
\label{ckt_1} 
\end{figure}
While all hermitian Hamiltonians can be realized by unitary operations, the
PT-symmetric Hamiltonian (being non-hermitian) can cause non-unitary evolution
and is hence nontrivial to simulate on a quantum processor.
The PT-symmetric Hamiltonian can be simulated by
enlarging the system into a higher-dimensional 
Hilbert space by using an ancillary qubit.
The PT-symmetric Hamiltonian for a 
single qubit can be written as ~\citep{Bender-iop-2007}:
\begin{equation}
\label{hamil}
H_{PT}=\sigma_{x}+ir\sigma_{z}
\end{equation}
where $r>0$ is the amount of non-hermiticity and
{$\sigma_{x},\sigma_{y},\sigma_{z}$ are Pauli matrices. The PT-symmetric
Hamiltonian satisfies the condition $(PT)H^{\dagger}(PT)^{-1}=H$, where
$P=\sigma_{x}$ is the parity operator and $T$ denotes the complex conjugation
operator. The energy gap $g$ of the PT-symmetric  Hamiltonian is
$g=2\sqrt{1-r^{2}}$ and its eigenvalues are $\pm\sqrt{1-r^{2}}$; for $\mid r
\mid<1$ the eigenvalues are positive, which means that the PT-symmetry is
unbroken and for $\mid r \mid>1$ the eigenvalues become complex, which leads to
a broken PT-symmetry. The Hamiltonian has an exceptional point at $\mid r
\mid=1$, where both the eigenvalues as well as the eigenvectors coalesce.

The evolution of $H_{PT}$ can be realized by introducing an ancillary qubit to
simulate the non-unitary operator $U_{PT}=e^{-iH_{PT}t}$, where $t$ represents
evolution time and the matrix form of the local operation
$U_{PT}=e^{-iH_{PT}t}$ is represented by 
\begin{equation}\label{unitary_PT}
U_{PT}=
\begin{bmatrix}
\cos\omega t-\frac{ir\sin\omega t}{\omega}&-\frac{i\sin\omega t}{\omega}\\
-\frac{i\sin\omega t}{\omega} &\cos\omega t+\frac{ir\sin\omega t}{\omega}
\end{bmatrix}
\end{equation}
where $\omega=\sqrt{1+r^{2}}$. Thus, this non-unitary operator can be
implemented by a linear combination of unitary operators using an ancillary
qubit~\citep{Zheng-epl-2018,Guilu-commTh_2006,Wen-pra-2019} and the quantum
circuit to implement it is shown in Fig.\ref{ckt_1}(a), where 
the initial state $\ket{\psi}\ket{0}$ is first 
created. Then the unitary
operator $V$ is performed on ancillary qubit which is
represented as \citep{Wen-pra-2019}:

\begin{equation}
V=\begin{bmatrix}
\cos{\theta}&-\sin{\theta}\\
\sin{\theta}&\cos{\theta}\\
\end{bmatrix}
\end{equation}
where
\begin{equation}\label{theta}
\cos\theta=\sqrt{\frac{1-r^{2}\cos^{2}{gt}/2}{1-r^{2}\cos{gt}}} \,
, \sin\theta=\frac{r\sin{gt/2}}{\sqrt{1-r^{2}\cos{gt}}}
\end{equation}
The two controlled unitary operations $C_{U_{1}}$ and $C_{U_{2}}$
are then performed on the first qubit of the two-qubit work state using
the ancillary qubit and the matrix form of these operations are represented
by~\citep{Wen-pra-2019}:

\begin{equation}
C_{U_{1}}=\begin{bmatrix}
U_{1}&0\\
0&I\\
\end{bmatrix}
\quad  \quad 
C_{U_{2}}=\begin{bmatrix}
I&0\\
0&U_{2}\\
\end{bmatrix}
\end{equation}
where
\begin{equation}
U_{1}=\begin{bmatrix}
\cos{\phi}&i\sin{\phi}\\
i\sin{\phi}&\cos{\phi}\\
\end{bmatrix}
\quad  \quad U_{2}=\sigma_{z}
\end{equation}
and 
\begin{equation}\label{phi}
\cos\phi=\frac{g\cos{gt/2}}{2\sqrt{1-r^{2}\cos^{2}{gt/2}}},
\sin\phi=\frac{-r\sin{gt/2}}{\sqrt{1-r^{2}\cos^{2}{gt/2}}}
\end{equation}
and the last operation is a Hadamard gate on 
the ancillary qubit. 
The state is then measured in the subspace 
spanned by $\vert 0 \rangle$
of 
the ancillary qubit, which leads to  the 
final state where the local PT-symmetric 
Hamiltonian has been implemented.

\begin{figure}[t]
\includegraphics[angle=0,scale=1]{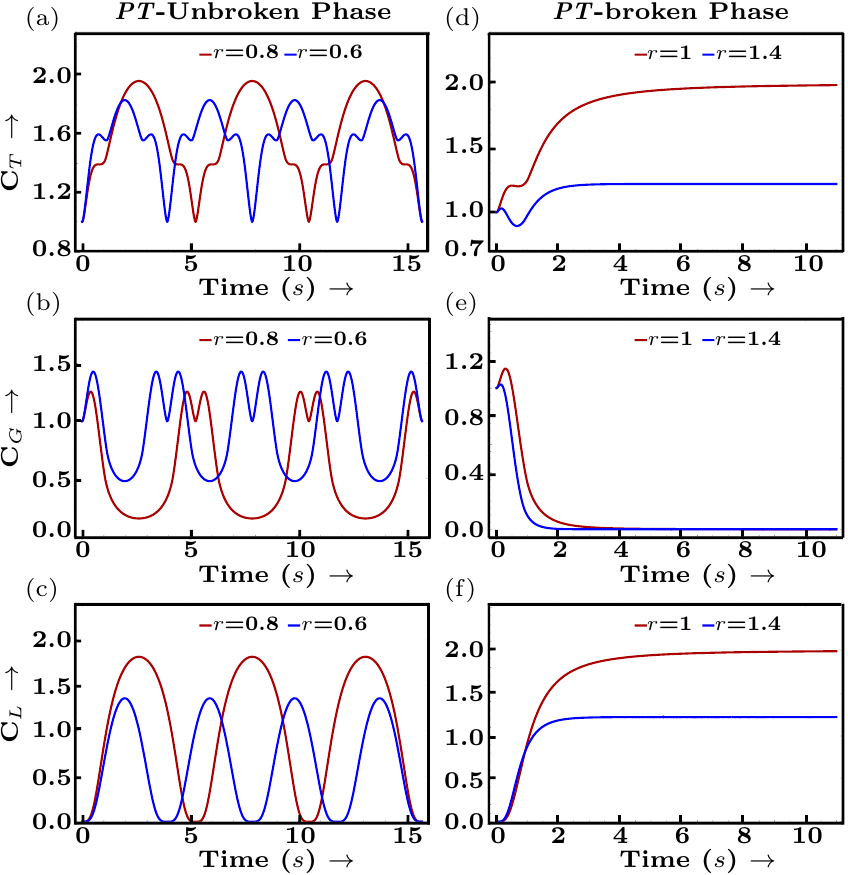} 
\caption{Plots of the evolution of
quantum coherence present in the maximally entangled bipartite (Bell) state
under different phases of PT-symmetry. (a), (b) and (c) represent the dynamics
of total coherence ($C_{T}$), global coherence ($C_{G}$) and local coherence
($C_{L}$) respectively, in the unbroken phase (at $r=0.6$ and $r=0.8$). (d), (e)
and (f) represent the dynamics of 
total, global and local coherence respectively,
in the broken phase (at $r=1.4$ and at exceptional point $r=1$).}
\label{bs1} 
\end{figure}

\section{Dynamics of quantum coherence under the PT-symmetric Hamiltonian}
\label{dyna}
\subsection{Measures of quantum coherence}
\begin{figure}[t]
\includegraphics[angle=0,scale=1]{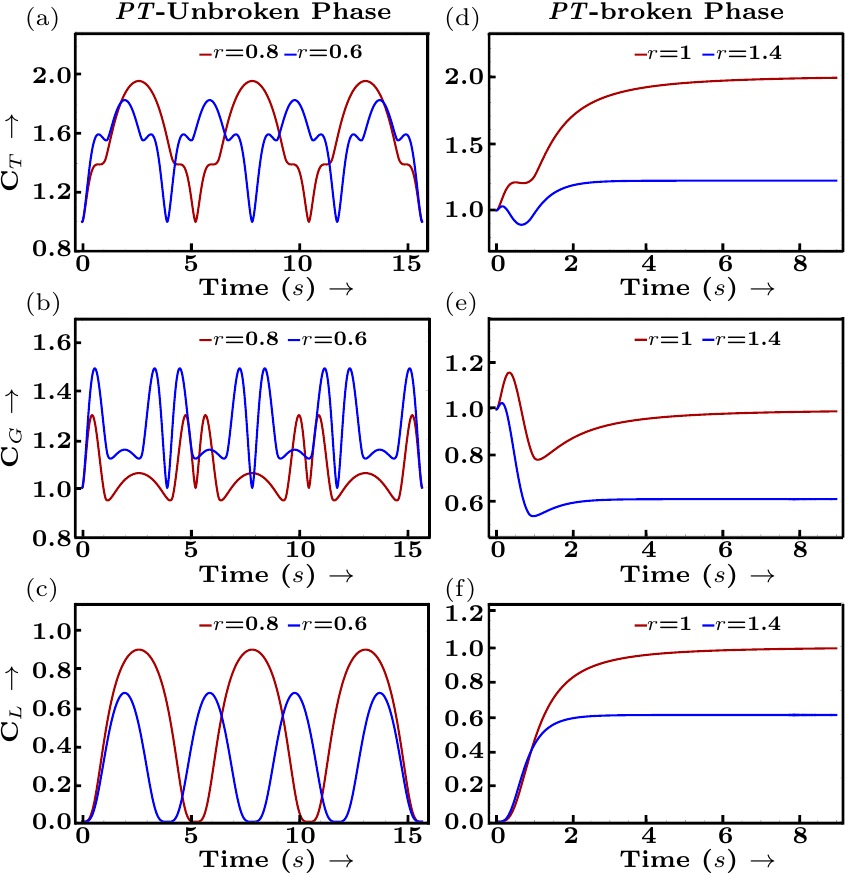}
\caption{Plots of the evolution of quantum coherence 
present in a maximally entangled tripartite
(GHZ) state under different phases of PT-symmetry. (a), (b) and (c) represent
the dynamics of total coherence ($C_{T}$), global coherence ($C_{G}$) and local
coherence ($C_{L}$) respectively, in the unbroken phase (at $r=0.6$ and
$r=0.8$). (d), (e) and (f) represent the dynamics of 
total, global and local coherence
respectively, in the broken phase (at $r=1.4$ and at exceptional point
$r=1$).} 
\label{ghz1} 
\end{figure}

Several measures have been proposed to quantify quantum
coherence that follow certain properties such as relative entropy 
and $l_{1}$ norm \citep{Baugratz-prl-2014}. These measures have
further been utilized to quantify different types of quantum coherences present
in multipartite systems such as total coherence, local coherence and global
coherence which are described below
~\citep{Cao-pra-2020,Radha-prl-2016}.

\textit{Total quantum coherence-} The total quantum
coherence~\citep{Baugratz-prl-2014} of the system is defined as the measure of
its distance to the closest incoherent state, where the incoherent state has the
form $\sigma=\sum_{i}p_{i}|i\rangle \langle i|$ and the fixed local basis
$\lbrace |i\rangle \rbrace$ for each qubit is $\lbrace|0\rangle,|1\rangle
\rbrace$. The relative entropy of coherence is defined as
~\citep{Baugratz-prl-2014}: 

\begin{equation}
C_{T}(\rho)=min_{\sigma\epsilon I}S(\rho\parallel\sigma)=S(\rho_{_{d}})-S(\rho)
\label{total_coh}
\end{equation}
where $S(\rho)=-tr(\rho log_{_{2}}\rho)$ is the Von-Neumann entropy 
of $\rho$, and $\rho_{_{d}}$ is the matrix of $\rho$ 
with all off-diagonal terms set to zero in the basis $|i\rangle$.

\textit{Local coherence-} Quantum coherence 
which is localized on each qubit of the entire system is called local
coherence~\citep{Cao-pra-2020,Radha-prl-2016}. For example, in a separable
two-qubit state
$|+-\rangle=(|0\rangle+|1\rangle)(|0\rangle-|1\rangle)/2$, quantum coherence
exists in each qubit. Local coherence can also be defined in terms of the
relative entropy ~\citep{Cao-pra-2020,Radha-prl-2016}:
\begin{equation}
C_{L}(\rho)=min_{\sigma\epsilon I}S(\delta(\rho)\parallel 
\sigma)=S(\delta_{d}(\rho))-S(\delta(\rho))
\label{local_coh}
\end{equation}
where $\delta(\rho)=\rho_{_{1}}\otimes\rho_{_{2}}$
($\delta(\rho)=\rho_{_{1}}\otimes\rho_{_{2}}\otimes\rho_{_{3}}$) for
a two-qubit state (three-qubit state) and
$\rho_{_{1}}=Tr_{_{2}}\rho_{_{12}}$ ($\rho_{_{1}}=Tr_{_{23}}\rho_{_{123}}$) is
the single-qubit reduced density matrix;  $\delta_{d}(\rho)$ is the matrix
produced by eliminating all off-diagonal elements of $\delta(\rho)$ in the
basis $|i\rangle$.

\textit{Global Coherence-} Quantum coherence that originates due to the
collective nature of the whole system is called global
coherence~\citep{Cao-pra-2020,Radha-prl-2016}. Mathematically, it is the
difference between total quantum coherence and local coherence
~\citep{Cao-pra-2020,Radha-prl-2016}:
\begin{equation}
C_{G}(\rho)=C_{T}(\rho)-C_{L}(\rho)
\label{global_coh}
\end{equation}
\subsection{Two-qubit maximally entangled (Bell) state}
Consider a two-qubit Bell state  defined as
\begin{eqnarray}
\rho_{_{BS}}=|\psi_{_{BS}}\rangle\langle\psi_{_{BS}}|\quad
{\rm with}\nonumber \\
|\psi_{_{BS}}\rangle=\frac{1}{\sqrt{2}}
\left(|00\rangle+|11\rangle\right) 
\label{bell-state}
\end{eqnarray}
Our aim is  to study the dynamics of various types of quantum coherences, namely
the total, global and  local coherence as the state evolves from $\rho_{BS}$
under the local operation generated by the PT-symmetric non-hermitian
Hamiltonian~\citep{Bender-iop-2007}( Eq.(\ref{hamil})), acting on one of the
qubits.  The final density matrix after the action of $U_{PT}^{2}=U_{PT}\otimes
I$ on the initial state $\rho_{BS}$ is given by: 
\begin{equation}
\label{densitybs}
\rho_{_{BS}}(t)=
\frac{U_{PT}^{2}\rho_{_{BS}}(0)U_{PT}^{2}}{Tr[U_{PT}^{2}
\rho_{_{BS}}(0)U_{PT}^{2}]}
\end{equation}
From this state, we calculate the total, global and local coherence by using the
definitions given in Eqs.~\ref{total_coh},\ref{local_coh} and \ref{global_coh},
respectively.  The regions of interest are the broken and unbroken phase regimes
of the PT-symmetric Hamiltonian and the exceptional
point~\citep{Bender-iop-2007}.

The results of these calculations  for the unbroken and broken PT-symmetry
regimes are plotted in Fig.~\ref{bs1}.  Figs.~\ref{bs1}(a)-(c) contain the plots
of the total, global and local coherence as a function of time for for the
unbroken PT-symmetry regime, while Figs.~\ref{bs1}(d)-(f) contain the same for
the broken PT-symmetry regime.  It can be observed from Fig.\ref{bs1}(a)-(c)
that all the different types of coherences (total, global and local coherence)
show an oscillatory behavior for the unbroken PT-symmetry regime. The amplitude
of the total and local coherence increases as the parameter $r$ approaches to
$r=1$ (exceptional point), while the amplitude of global coherence decreases as
the parameter $r$ approaches the exceptional point.  The transition in the
dynamics of quantum coherence once the exceptional point is reached, can be
observed in Figs.\ref{bs1}(d)-(f), where the amplitude of total quantum
coherence initially increases and then freezes at later times.  A similar
behavior is observed in the dynamics of local coherence in the broken phase
regime (which is initially absent and is created when acted upon by the
PT-symmetric Hamiltonian) where its amplitude first increases and then attains a
stable value.  The freezing behavior of quantum coherence depicted in
Fig.~\ref{bs1} is different from the dynamics of previously observed quantum
correlations (both entanglement and quantum discord) in the maximally entangled
state in the broken phase.  
The global coherence in the broken phase of PT-symmetry
first increases with time and then decays exponentially to zero.
The discussion of the dynamics of various
coherences at the exceptional point ($r=1$) will be taken up 
later in Section~\ref{exp-pt}.

\begin{figure}[t]
\includegraphics[angle=0,scale=1]{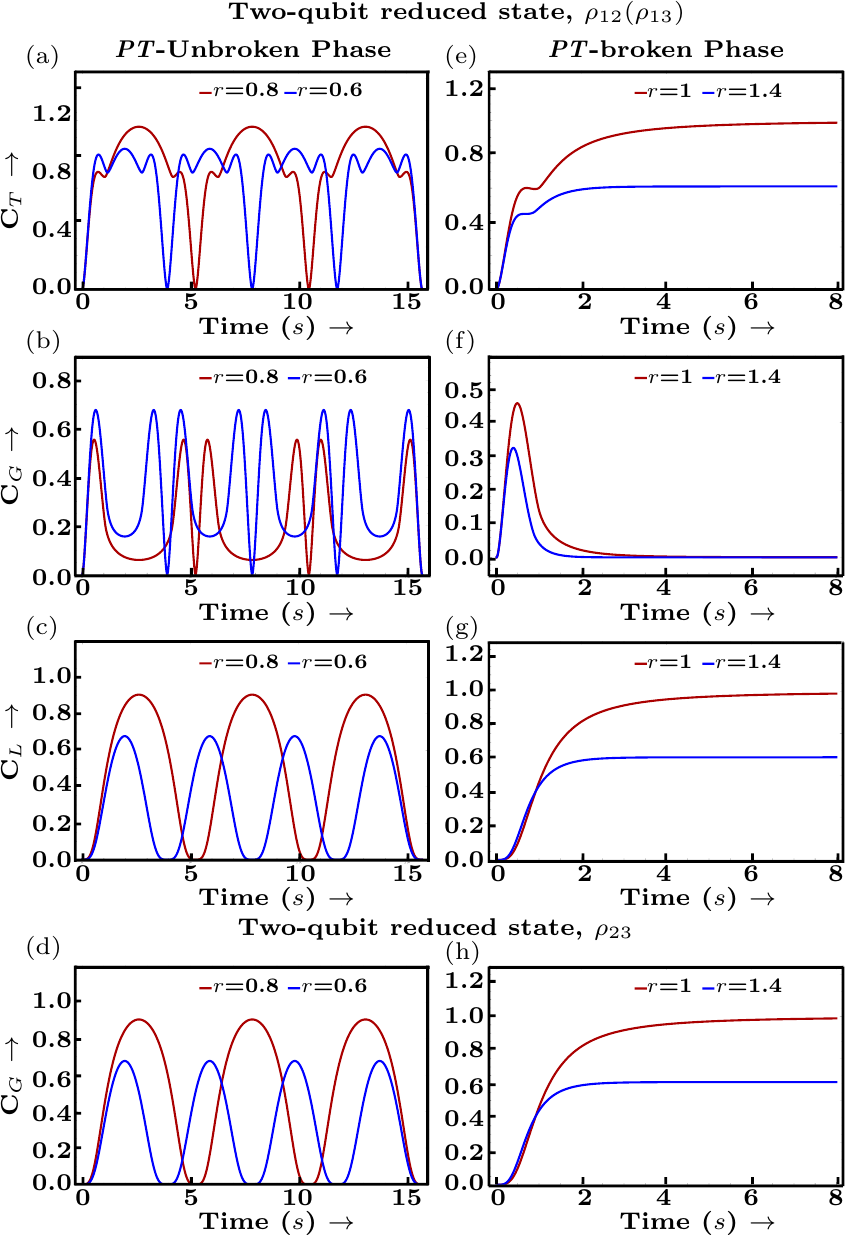}
\caption{Plots of the evolution of quantum coherence present
in the two-qubit reduced GHZ state  under different phases
of PT-symmetry. (a), (b) and (c) represent the dynamics of
total coherence ($C_{T}$), global coherence ($C_{G}$) and
local coherence ($C_{L}$) respectively, present in
$\rho_{12}(\rho_{13})$ in unbroken PT-symmetry phase
($r=0.6$ and $r=0.8$). (e),(f) and (g) represent the
dynamics of total, global and local coherence respectively,
present in $\rho_{12}(\rho_{13})$ in a broken phase $r=1.4$
and at the exceptional point $r=1$.  (d) represents the
dynamics of global coherence present in $\rho_{23}$ in an
unbroken and (h) represents the dynamics of global coherence
present in $\rho_{23}$ in the broken phase of PT-symmetry.} 
\label{ghz2} 
\end{figure}
\begin{figure*}[t]
\includegraphics[angle=0,scale=1]{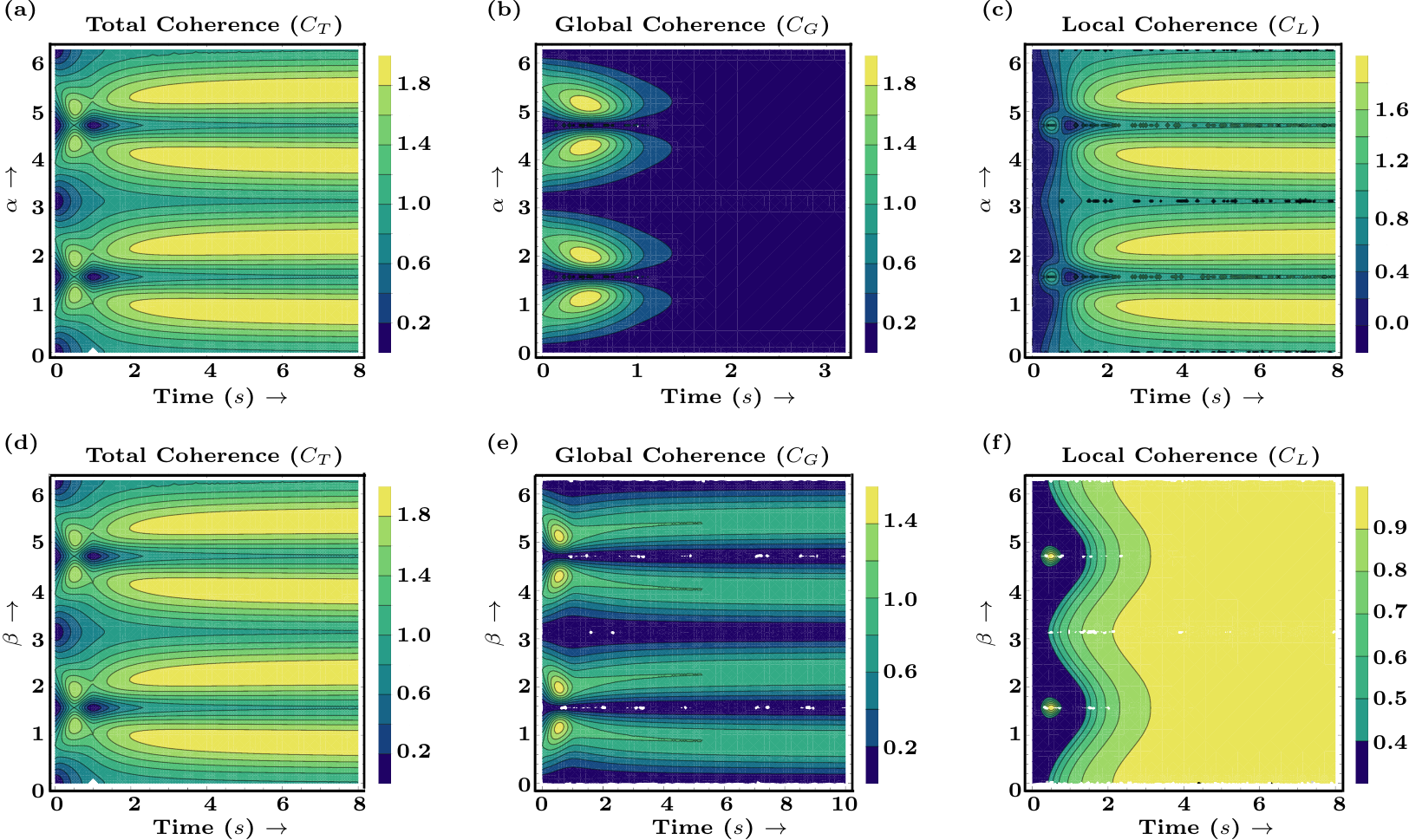}
\caption{
Contour plots of the dynamics of 
(a) total coherence, (b) global coherence, and (c) local coherence, 
present in
the state 
$|\psi_{_{BS}}^{\alpha}\rangle$ at the exceptional point.
Contour plots of  the dynamics of 
(d) total coherence, (e) global coherence, and (f) local coherence, 
present in the state
$|\psi_{_{GHZ}}^{\beta}\rangle$ 
at the exceptional point; the angles $\alpha$ and
$\beta$ vary from $0$ to
$2\pi$ (6.28) radians. }
\label{cont1} 
\end{figure*} 
\subsection{Three-qubit maximally entangled state (GHZ state)}
Next we consider three-qubit states and analyze the
evolution of their coherence properties when one of the
qubits is subjected to time  evolution generated by the
PT-symmetric Hamiltonian (Eq.~(\ref{hamil})).
We first consider the maximally entangled 
Greenberger-Horne-Zeilinger (GHZ) state:
\begin{eqnarray}
\rho_{_{GHZ}}=|\psi_{_{GHZ}}\rangle
\langle\psi_{_{GHZ}}|\quad {\rm where} \nonumber \\
|\psi_{_{GHZ}}\rangle=\frac{1}{\sqrt{2}}(|000\rangle+|111\rangle)
\end{eqnarray}
Taking the GHZ state as the initial state, we consider the action of
the PT-symmetric Hamiltonian (Eq.~(\ref{hamil})) on
the first qubit. This amounts to evaluating the action of the operator
$U_{PT}$ defined in Eq.~(\ref{unitary_PT}) on the first
qubit, leading to the
overall operation on the three-qubit system to be
implemented as: 
\begin{equation}
U_{PT}^{3}=U_{PT}\otimes I \otimes I.
\end{equation}
The final normalized density operator corresponding to the local action of
the PT-symmetric Hamiltonian on the GHZ state then can be written
as: 
\begin{equation}\label{densitygs}
\rho_{_{GHZ}}'(t)=\frac{U_{PT}^{3}
\rho_{_{GHZ}}(0)U_{PT}^{3}}{Tr[U_{PT}^{3}\rho_{_{GHZ}}(0)U_{PT}^{3}]}
\end{equation}
From this final state, the three coherences under consideration can be
calculated as per the definitions given in Eqs.~\ref{total_coh},\ref{local_coh}
and \ref{global_coh}, and their dynamics can be studied as was done for the case
of the two-qubit Bell state in the previous section.

The results of these calculations for unbroken and broken PT-symmetry regimes
are presented in Figure~\ref{ghz1}.  As can be seen from
Figs.~\ref{ghz1}(a)-(c),  in the unbroken PT-symmetry phase, all quantum
coherences in tripartite state show an oscillatory behavior similar to the
bipartite state.  As the parameter $r$ increases and reaches the exceptional
point, the transition in dynamics of total and local coherence in tripartite
state can be observed, wherein the quantum coherence freezes and attains a
stable value at later times.  On the other hand, the dynamics of global
coherence in the GHZ state is different from the Bell  state, as can be
observed from Fig.\ref{ghz1}(c), where the global coherence initially decays,
then increases and finally freezes at later times.

Next, the dynamics of quantum coherences in three possible two-qubit reduced
states ($\rho_{ij}=tr_{k}\left(\rho_{ijk} \right)$, where $i,j,k \in (1,2,3)$)
of the GHZ state are studied.  The results of the calculations of dynamics of
various coherences for the two-qubit reduced states associated with the GHZ
state are presented in Figure~\ref{ghz2}.  It is observed that various quantum
coherence (total, local and global) are generated in two-qubit reduced states,
$\rho_{12}$ and $\rho_{13}$ due to the local operation of the PT-symmetric
Hamiltonian on the first qubit in both unbroken and broken phase.  The pattern
of quantum coherence evolution in the two-qubit reduced states ($\rho_{12}$ and
$\rho_{13}$) are similar to the bipartite maximally entangled Bell state as
shown in Fig.\ref{ghz2}(a)-(c) and Fig.\ref{ghz2}(e)-(g).  However in the
two-qubit reduced state $\rho_{23}$, only global coherence is generated, which
contributes to overall coherence.  From Figs.\ref{ghz2}(d) and (h) it can be
seen that the dynamics of global coherence present in $\rho_{23}$ shows an
oscillatory behavior in unbroken phase while in broken phase it initially
increases and then freezes at later times. 
\subsection{Dynamics of coherence at the exceptional point}
\label{exp-pt}
The results of the calculations  at the exceptional point ($r=1$) for the
two-qubit Bell state and the three-qubit GHZ state are depicted as contour plots
in Fig.~\ref{cont1}.  The dynamics of various quantum coherences present in
$|\psi_{_{BS}}^{\alpha}\rangle= \cos\alpha|00\rangle+\sin\alpha|11\rangle$ were
explored at the exceptional point  where $\alpha$ is varied from $0$ to $2\pi$
\ie the state evolves from the ground state to an excited state, passing through
the maximally entangled state. From the contour plots shown in
Fig.\ref{cont1}(a)-(c), it can be observed that the total coherence stabilizes
with time but the amplitude of total coherence varies with $\alpha$. Similar
dynamics of local coherence can be observed from the contour plot, where it can
be seen that the local coherence decays with time and does not depend on the
$\alpha$ value.  The amplitude of the global coherence at the exceptional point
first increases with time and then decays exponentially to zero.  The dynamics
of different quantum coherences in the broken phase are similar to the dynamics
of quantum coherence observed at the exceptional point except that their
amplitude decreases as the parameter $r$ increases.

The dynamics of various types of quantum coherences present in
$|\psi_{_{GHZ}}^{\beta}\rangle=\cos\beta|000\rangle+\sin\beta|111\rangle$ were
investigated at the exceptional point.  The parameter $\beta$ is varied from $0$
to $2\pi$, \ie the state goes  from the ground state $|000\rangle$ to the
excited state $|111\rangle$, through the maximally entangled state.  The contour
plots of the dynamics of the quantum coherences are shown in
Figs.\ref{cont1}(d)-(f), from where it can be observed that the total coherence
freezes with time and the amplitude of the frozen coherence varies with varying
$\beta$ values of the state. Similar dynamics of global coherence can be
observed from the contour plot, which shows that the global coherence freezes
with time and its amplitude depends on the $\beta$ value. The contour plot of
the local coherence depicts that  the  local coherence stabilizes to a value $1$
with time and does not depend on the $\beta$ value.  The dynamics of all quantum
coherences in the broken phase are similar to the dynamics of the quantum
coherences observed at the exceptional point but their amplitude decreases as
the parameter $r$ increases as displayed in Figure~\ref{ghz1}(d)-(f) where panel
(d) shows evolution of total coherence, panel (e) shows evolution of the global
coherence and  evolution of the local coherence is shown in panel (f).

\section{Experimental demonstration of the dynamics of 
quantum coherence under a PT-symmetric Hamiltonian}
\label{Exp}
As discussed in Sec.\ref{simu}, the PT-symmetric Hamiltonian
can be simulated by the dilation method using an ancillary
qubit. The experimental implementation of PT-symmetric
Hamiltonian on an NMR quantum processor is realized using a
three-qubit system where the three spin-1/2 nuclei ($^{1}$H,
$^{19}$F and $^{13}$C) of $^{13}$C-labeled
diethylfluoromalonate dissolved in acetone-D6, are encoded
as the first, second and third qubit respectively; the
$^{1}$H and $^{19}$F are used as a two-qubit system while
the $^{13}$C spin is utilized as the ancillary qubit.

\begin{figure}[h]
\includegraphics[angle=0,scale=1]{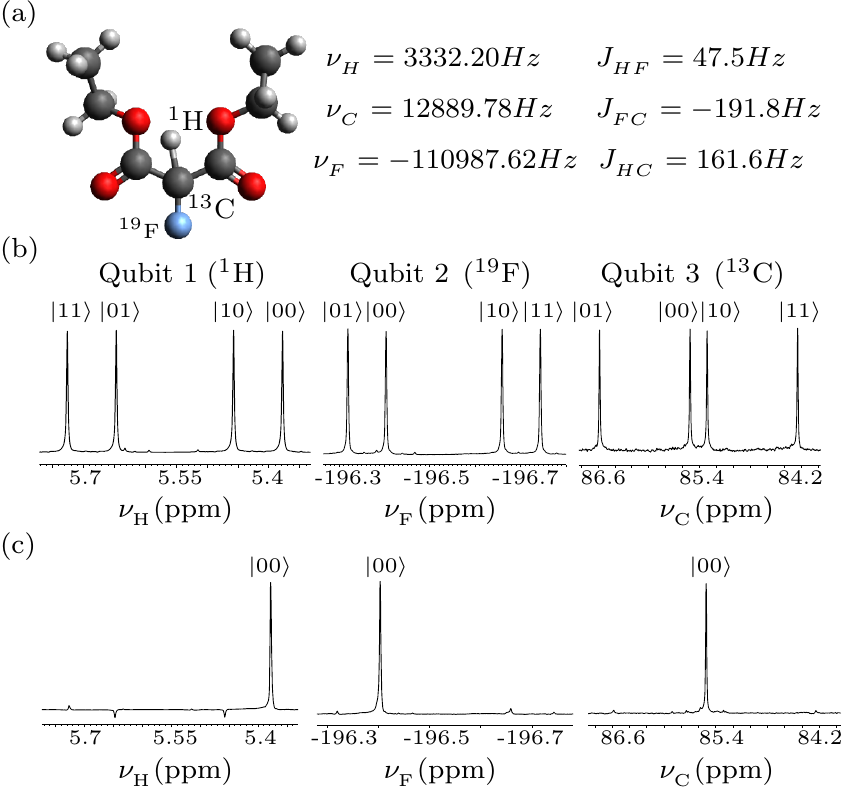}
\caption{(a) Molecular structure of $^{13}$C-labeled
diethylfluoromalonate with the three qubits encoded as
$^{1}$H, $^{19}$F and $^{13}$C and the system parameters
given alongside.  (b) NMR spectrum of thermal equilibrium
state obtained after a $\pi/2$ readout pulse. (c) NMR
spectrum of the pseudopure state. Each resonance peak is
labeled in the logical basis and the horizontal scale of
each spectrum denotes the chemical shift in ppm.}
\label{mol} 
\end{figure}

The molecular structure of the three-qubit system along with
tabulated system parameters and the NMR spectrum of each
qubit at thermal equilibrium are shown in Fig.\ref{mol}(a)
and Fig.\ref{mol}(b), respectively. The NMR Hamiltonian for
a three-qubit system in a rotating frame under the weak
approximation ~\citep{ernst-book-90} is given by

\begin{equation}\label{NMR_Hamilt}
H=-\sum_{i=1}^{3}\nu_{i}I_{z}^{i}+\sum_{i>j,i=1}^{3}J_{ij}I_{z}^{i}I_{z}^{j}
\end{equation}
where $i,j=$1, 2 and 3 label the nuclear spin, $v_{i}$ is the chemical
shift of the $i$th spin, $J_{ij}$ represents the scalar spin-spin coupling
strength and $I_{z}^{i}$ is the $z$ component of the spin angular momentum
operator for the $i^{th}$ spin. The three-qubit system was initialized into
a pseudopure state (PPS) $|000\rangle$ using the spatial
averaging technique ~\cite{cory-physD-98,mitra-jmr-07}, with the density
$\rho_{000}$ given by
\begin{equation}\label{PPS}
\rho_{000}=\frac{(1-\epsilon)}{8}I +\epsilon
\vert 000 \rangle \langle 000 \vert
\end{equation}
where $\epsilon\sim 10^{-5}$ represents the spin polarization at room
temperature and $I$ is the 8$\times$8 identity operator. The NMR spectrum of
the three-qubit PPS is shown in Fig.\ref{mol}(c). The experimentally prepared
PPS was tomographed and its fidelity was computed to be $0.9857\pm0.0001$ using
the Uhlmann-Jozsa fidelity measure \citep{uhlmann-rpmp-76,jozsa-jmo-94}.

The density matrix of the initial state was reconstructed using least square
constrained convex optimization method~\citep{Gaikwad-qip-2021} by performing
full quantum state tomography using a set of seven preparatory pulses $\lbrace
III$, $XXX$, $IIY$, $XYX$, $YII$, $XXY$, $IYY \rbrace$ where $I$ represents
'no-operation' and $X(Y)$ is the local $\pi/2$ unitary operator with phases
$x(y)$ and local operations are achieved using spin-selective $\pi/2$
radiofrequency (rf) pulses. All experiments were performed at room temperature
293K on a Bruker Avance-III 600 MHz FT-NMR spectrometer equipped with a QXI
probe.

\begin{figure}[H]
\includegraphics[angle=0,scale=1]{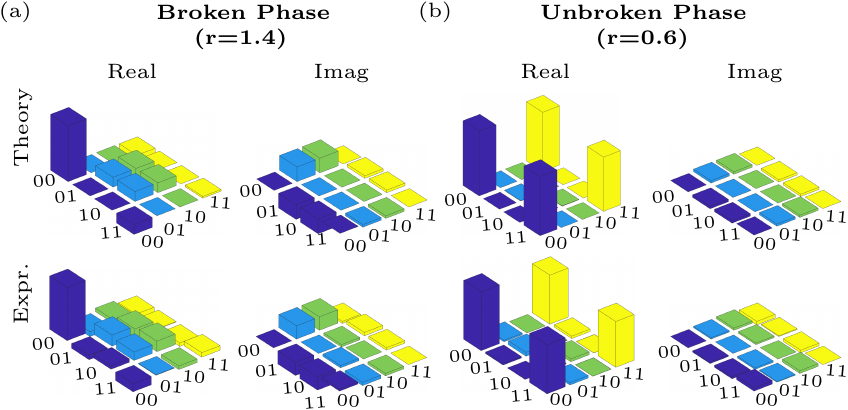}
\caption{Real (left) and imaginary (right) parts of 
the theoretically expected and
experimentally reconstructed density matrix of 
the Bell State after implementation of
the PT-symmetric Hamiltonian at time $t=4s$ under (a) a broken phase at
non-hermiticity strength r=1.4 and (b) an unbroken phase at
non-hermiticity strength r=0.6.}
\label{tomo} 
\end{figure}

The quantum circuit to implement PT-symmetric Hamiltonian is shown in
Fig.\ref{ckt_1}(a). The first block creates the maximally entangled
bipartite state (Bell State, $\rho_{_{BS}} \otimes \rho_{0}$) on two qubits
($\rho_{_{BS}}$) and the third qubit ($\rho_{0}$) is 
utilized as an ancillary qubit.
The next block of the circuit implements a PT-symmetric Hamiltonian on the
first qubit of the two-qubit state using an ancillary qubit where the parameter
$r$, and the time period ($t$) are varied by $\theta$ and $\phi$ as given in
Eq.~[\ref{theta}] and Eq.~[\ref{phi}] respectively; $\theta$ is controlled
by the operator $V$ and $\phi$ by the operator controlled-$U_{1}$. All quantum
gates were implemented using rf-pulses and free evolution under the system
Hamiltonian. Local rotations were implemented by spin-selective pulses
with different phases; the durations of $\pi/2$ pulses for $^{1}$H,
$^{19}$F and $^{13}$C nuclei were 9.38 $\mu$s, 21.8 $\mu$ s and 15.1 $\mu$ s
at power levels of 18.14 W, 42.27 W and 179.47 W, respectively.

\begin{figure*}[t]
\includegraphics[angle=0,scale=1]{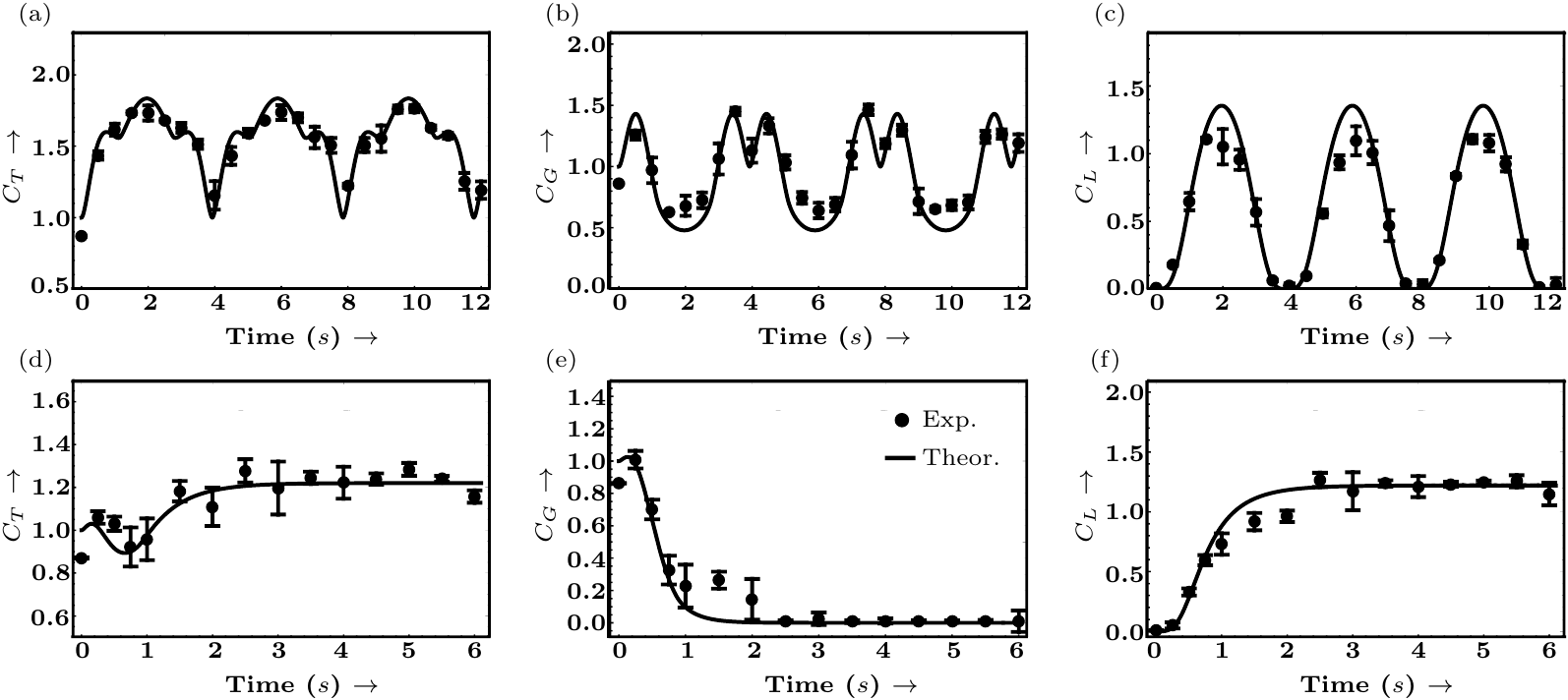}
\caption{Dynamics of quantum coherence present in the two-qubit 
maximally entangled Bell State under a
$PT$-Symmetric Hamiltonian. Plots in (a), (b) and (c) represent the dynamics of
the total coherence ($C_{T}$), global coherence ($C_{G}$) 
and local coherence
($C_{L}$) respectively, in the unbroken phase of PT-Symmetry ($r=0.6$).
(d), (e) and (f) represent the dynamics of 
the total coherence ($C_{T}$), global
coherence ($C_{G}$) and local coherence ($C_{L}$) respectively, in the
broken phase of PT-Symmetry ($r=1.4$). The black lines are 
the theoretical
curves (Theor.) and the black filled circles are the 
experimental points (Exp.).}
\label{res} 
\end{figure*}

The NMR pulse sequence used to experimentally prepare 
a maximally entangled bipartite state 
is shown in Fig.\ref{ckt_1}(b).
The fidelity of the initial state $\left(\rho_{_{BS}} \otimes \rho_{0}\right)$
created was found to be $0.9745\pm0.0102$.  The method to simulate the
PT-symmetric Hamiltonian is described in Sec.\ref{simu}, where $\theta$ and
$\phi$ control the degree of non-hermiticity $r$ and the time point of evolution
$t$. We experimentally simulated the PT-symmetric Hamiltonian  by choosing
parameters $r=0.6$ and $r=1.4$ corresponding to unbroken and broken phase
respectively, which was implemented on the first qubit of the bipartite state,
using the third qubit as an ancillary qubit. The parameter $r$ was fixed
depending on different phases of PT-symmetry, and the state was allowed to
evolve by increasing the time period $t$ and accordingly varying $\theta$ and
$\phi$. Each state was tomographed at different time points after implementing
the PT-symmetric Hamiltonian in both the unbroken and broken phase.  The final
density matrices of the two qubit-Bell states were reconstructed by measuring in
the $|0 \rangle$ subspace of the ancillary qubit, where only three tomographic
pulses ($III$, $IIY$, $YII$) were required to reconstruct the entire density
matrix. A representative tomograph of the experimentally reconstructed density
matrix of the Bell state after the implementation of the PT-symmetric
Hamiltonian in two different phases (unbroken phase at $r=0.6$ and broken phase
at $r=1.4$) of PT-symmetry at time $t=4$~s is shown in Fig.\ref{tomo}. 

A single qubit coherence flow in PT-symmetric and
anti-PT-symmetric systems was demonstrated experimentally using an optical
setup and it was
observed that coherence flow oscillates back and forth in the 
unbroken 
PT-symmetric and anti-PT-symmetric regimes while in broken phase,
coherence flow attains a stable value in both PT-symmetric and anti-symmetric
systems~\cite{Fang-commphy-2021}. 
We investigated the dynamics of quantum coherence (total, local and
global coherence) 
in the  maximally entangled (Bell) state, when one qubit of the Bell
state is acted upon by the PT-symmetric Hamiltonian at $r=0.6$ (unbroken phase)
and at $r=1.4$ (broken phase). The results are plotted in Fig.\ref{res} where
(a), (b) and (c) depict the dynamics of various types of quantum coherences in
the unbroken phase and (d), (e) and (f) depict the dynamics of various types of
quantum coherence in the broken phase. 
Similar patterns of coherence dynamics are observed as in the case
of the optics setup~\cite{Fang-commphy-2021}.
Our results in
Fig.\ref{res} (a)-(c) show that in the Bell state, the total, local and
global coherences oscillate with time in the unbroken phase ($r=0.6$)
while the amplitudes are different for various quantum
coherences. Another interesting feature of local coherence generation is
observed that it oscillates with time, and is not present in the initial
Bell state. In the broken phase ($r=1.4$), the total coherence initially
increases and then freezes at later times. Similarly, the results of
Fig.\ref{res}(c) indicate that the local coherence, which is
not present initially in the Bell state, is created and increases under the
broken phase of PT-symmetry and freezes at later times.
Fig.\ref{res} (b) shows that the the global coherence initially
increases and then exponentially decays with time. Our experimental results
match well with the theoretically predicted behavior of quantum coherence,
within experimental errors. 

Similar patterns of information flow 
have been observed in NMR in a
single-qubit system 
where information flow oscillates in the broken phase while it
decays with time in the unbroken phase 
of anti-PT-symmetry~\cite{Wen-pra-2019}. 
A similar freezing pattern of quantum coherence 
has also been observed 
 in a two-qubit subsystem of a three-qubit
GHZ state under the PT-symmetric Hamiltonian using nuclear spins,
where it was observed that entropy and entanglement  
stabilize to a certain value~\cite{Wen-npj-2020}.

\section{Conclusion} 
\label{Con}
We have theoretically investigated the dynamics of quantum coherence (total,
global and local coherence) present in maximally entangled bipartite and
tripartite states in both the unbroken and the broken phase of PT-symmetry and
at an exceptional point when one qubit is acted upon by the local PT-symmetric
Hamiltonian. Our results show that the different types of quantum coherence
oscillate in the unbroken phase of PT-symmetry. In the broken phase of
PT-symmetry and at the exceptional point, the total and local coherence first
increase and then attain a stable value over time in both the maximally
entangled bipartite and tripartite states.  The global coherence decays
exponentially in the bipartite state but freezes over time in the tripartite
state. Our results indicate that quantum coherence behaves differently under
PT-symmetry, when the dimensionality of the quantum system is increased. The
dynamics of quantum coherence present in the maximally entangled bipartite
state was experimentally verified by implementing the PT-symmetric Hamiltonian
on an NMR quantum processor. Our work sheds some light on the effect of the
PT-symmetric Hamiltonian on quantum coherences in a multipartite system. Our
results can help in gaining an understanding of the effects of the PT-symmetric
Hamiltonian in quantum thermodynamics and in quantum information
processing}\citep{Wei-pre-2018,Deffner-prl-2015}.
\begin{acknowledgments}
All the experiments were performed on a Bruker Avance-III 600 MHz FT-NMR
spectrometer at the NMR Research Facility of IISER Mohali.  A.  acknowledges
financial support from DST/ICPS/QuST/Theme-1/2019/Q-68.  K~.D. acknowledges
financial support from DST/ICPS/QuST/Theme-2/2019/Q-74.
\end{acknowledgments}


%
\end{document}